\documentclass{article}

\PassOptionsToPackage{numbers,sort&compress}{natbib}

 \usepackage[main, final]{neurips_2025}

\usepackage[utf8]{inputenc} 
\usepackage[T1]{fontenc}    
\usepackage{hyperref}       
\usepackage{url}            
\usepackage{booktabs}       
\usepackage{amsfonts}       
\usepackage{nicefrac}       
\usepackage{microtype}      
\usepackage{xcolor}         
\usepackage{ulem}       
\usepackage{amsmath}
\usepackage{placeins}
\usepackage{bbding}
\usepackage{float}
\usepackage{makecell}
\usepackage[most]{tcolorbox}
\usepackage{lmodern}
\usepackage{textcomp}
\usepackage{placeins} 
\usepackage{setspace}

\usepackage{algorithm}
\usepackage{algorithmic}
\usepackage{tikz}
\usepackage{enumitem}
 \usepackage{arydshln}

\usepackage{multirow,tabularx} 

\newcommand*\circled[1]{\tikz[baseline=(char.base)]{
    \node[shape=circle,draw,inner sep=0.5pt] (char) {\scriptsize #1};}}

\newcommand{\modelname}{SegTune}

\title{SegTune: Structured and Fine-Grained Control for Song Generation}

\author{%
  Pengfei Cai\thanks{Equal contribution. Work done during internship at Kuaishou Technology.}, Joanna Wang\footnotemark[1], Haorui Zheng\footnotemark[1], Xu Li\thanks{Corresponding author: Xu Li (lixu15@kuaishou.com)}, Zihao Ji, \\[0.8ex]
  \textbf{Teng Ma}, \textbf{Zhongliang Liu}, \textbf{Chen Zhang}, \textbf{Pengfei Wan} \\[1.2ex]
  Kling Team, Kuaishou Technology
}

\begin{document}

\maketitle

\begin{abstract}
Recent advancements in song generation have shown promising results in generating songs from lyrics and/or global text prompts. 
However, most existing systems lack the ability to model the temporally varying attributes of songs, limiting fine-grained control over musical structure and dynamics. 
In this paper, we propose \textbf{SegTune}, a non-autoregressive framework for structured and controllable song generation. SegTune enables segment-level control by allowing users or large language models to specify local musical descriptions aligned to song sections.
The segmental prompts are injected into the model by temporally broadcasting them to corresponding time windows, while global prompts influence the whole song to ensure stylistic coherence. 
To obtain accurate segment durations and enable precise lyric-to-music alignment, we introduce an LLM-based duration predictor that autoregressively generates sentence-level timestamped lyrics in LRC format. We further construct a large-scale data pipeline for collecting high-quality songs with aligned lyrics and prompts, and propose new evaluation metrics to assess segment-level alignment and vocal attribute consistency. 
Experimental results show that SegTune achieves superior controllability and musical coherence compared to existing baselines. See \url{https://cai525.github.io/SegTune_demo/} for demos of our work.
\end{abstract}

\section{INTRODUCTION}

Song serves as a profound medium for emotional expression, carrying both cultural memory and personal sentiment.  
Recent advances in neural song generation have led to significant breakthroughs.  
Commercial systems  represented by Suno can synthesize complete songs including vocal performances and accompaniment within seconds via end-to-end modeling.  
The outputs are approaching professional production quality and are rapidly transforming the conventional music creation pipeline.
These advances not only lower the entry barrier for music creation—democratizing access to high-quality song production—but also offer new tools for professional musicians, fostering creativity and accelerating music production workflows.

Among various music related tasks, song generation  is one of the most challenging, which involves creating full songs with vocals, accompaniment, from lyrics and control signals.  
While commercial systems have demonstrated expert-level performance, the open-source community has also made considerable progress.  
These models can be broadly categorized into three paradigms: autoregressive language models~\cite{yue}, diffusion-based non-autoregressive models~\cite{acestepstep, diffrhythm,diffrhythm_plus,liu2025jamtinyflowbasedsong}, and hybrid systems~\cite{songcreator,liu2025songgen,music_cot,lei2025levo} that combine both architectures.

Most of these models rely on global control signals to guide song generation.
Control signals, typically provided as textual descriptions or tags, define overall musical characteristics such as genre, emotional tone, and vocal timbre.
While effective for control the overall style, relying on global prompts alone introduces several limitations.
One major challenge lies in the temporal dynamics of music.
Attributes such as instrumentation and emotion naturally vary across different segments of a song, while a single global description is insufficient to reflect expressive shifts throughout the piece.
Another limitation stems from the inherent complexity of the song generation task.
Unlike singing voice synthesis (SVS) systems that focus solely on vocal rendering with pre-defined musical notes, song generation models must handle both the composition of musical content and the rendering of audio in an end-to-end manner.
This requirement places significant strain on the model’s capacity, especially for non-autoregressive architectures.
Moreover,professional music production often demands precise control over specific song segments to align with creative intent.
For example, enhancing emotional intensity in the chorus while layering additional instrumentation such as guitar accompaniment is a common artistic intent.
Such localized control cannot be achieved through the global prompt alone.

\begin{figure}[t]
  \centering
  \includegraphics[width=\linewidth]{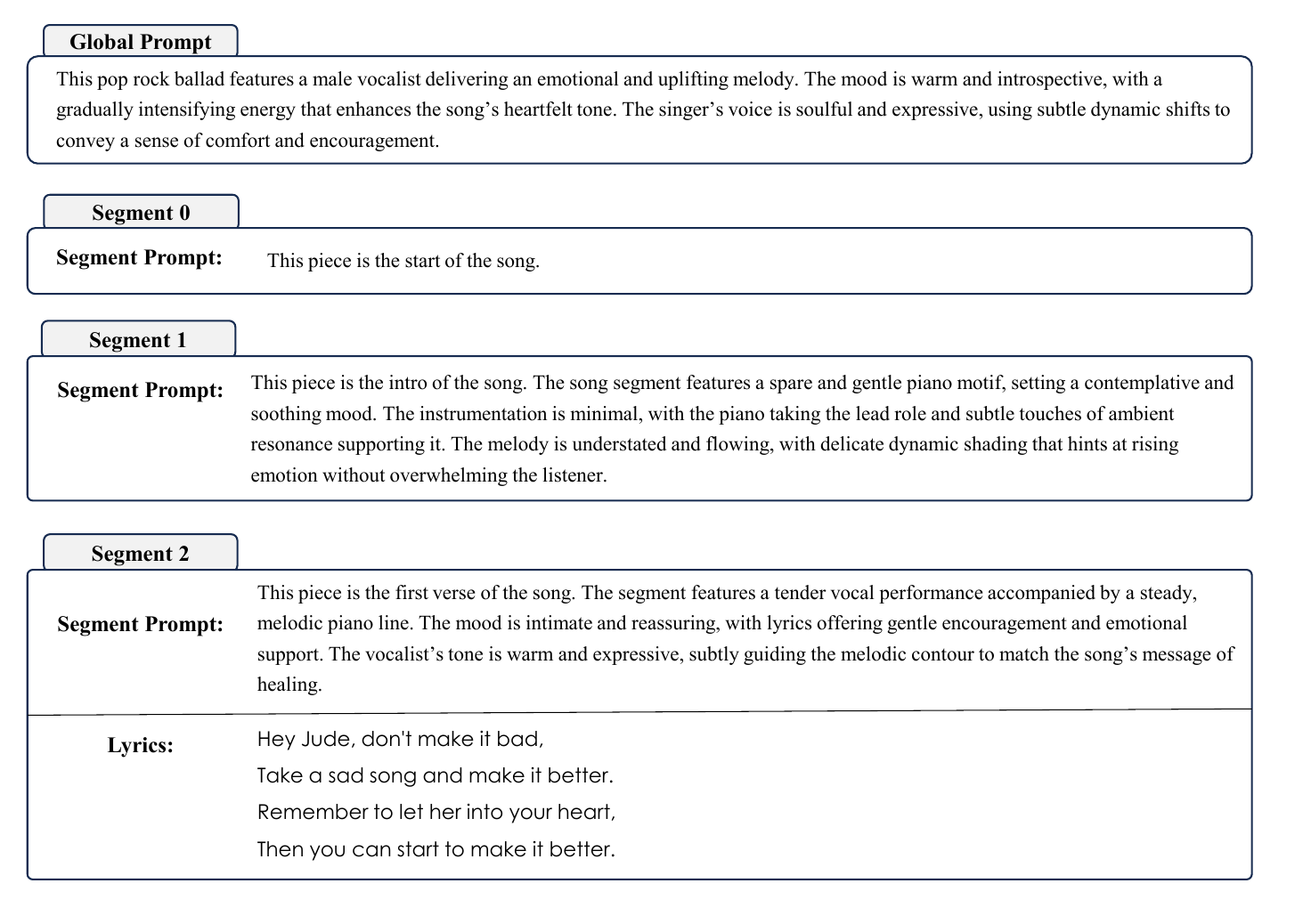}
  \vspace{-1.5em}
  \caption{The figure illustrates the input format of \modelname{}, which includes lyrics and textual prompts.  
  The global prompt describes overall musical attributes such as genre, vocal timbre, and global mood tune.  
  Segment prompts specify the attributes of individual segments, such as structure labels,  emotion change, or instrumentation. 
   Segment prompts  can either be manually provided by the user or automatically generated by the LLM during the prompt engineering stage.}
   \vspace{-1em}
  \label{fig:input}
\end{figure}

In this work, we introduce \textbf{\modelname{}}, a non-autoregressive (NAR) model for song generation task.
\modelname{} enables controllability at both the global and segment levels, as illustrated in Figure~\ref{fig:input}.  
Global prompts are provided by the user, while segment-level prompts can either be specified manually or generated automatically by a large language model (LLM) during the prompt engineering stage, which satisfys both professional music creators and amateur users.
To encode these multi-level prompts, we design a dedicated prompt encoder that injects segment-level information into its corresponding temporal window, while applying the global prompt across the entire sequence to maintain stylistic coherence.  
To accurately determine segment boundaries and enable sentence-level lyric alignment, we incorporate a fine-tuned LLM to predict sentence-level durations.  
This approach addresses a known limitation of previous NAR song generation models, which required manually specified durations, and leads to significant improvements in the musicality of the generated songs.

Our main contributions can be summarized as follows:

\begin{itemize}[leftmargin=0em]
    \item \textbf{Segment-level control for song generation.}  
    We propose a strategy that enables fine-grained control over musical attributes—such as emotion, rhythm, and instrumentation—at the segment level through textual prompts.
    
    \item \textbf{LLM-based duration prediction.}  
    We fine-tune a large language model to serve as a duration predictor, which autoregressively generates LRC-style lyrics with sentence-level timestamps conditioned on the lyrics and both global and local descriptions.
    
    \item \textbf{A comprehensive song data pipeline.}  
    We build a complete data processing pipeline to clean, organize, and annotate high-quality songs, along with cleaned lyrics and multi-level textual descriptions.

    \item \textbf{A systematic evaluation framework.}  
    In addition to conventional song generation metrics, we introduce new evaluation criteria to evaluate new features of our model.  
    A segment-level Mulan similarity score is used to assess the model's ability to follow segment-level instructions, and a singer attribute following score based on audio LLM is proposed to evaluate instruct following of  vocal attributes, addressing limitations in prior evaluation schemes.
\end{itemize}

\section{RELATED WORK}
\subsection{Music Generation}

Music generation focuses on producing coherent and stylistically consistent audio conditioned on text, melody, or other high-level cues. 
One prevalent approach is autoregressive modeling, which sequentially generates discrete audio representations. 
MusicGen~\cite{musicgen} combines residual vector quantization (RVQ) with Transformer-based decoder to improve fidelity and controllability, while MusicLM~\cite{musiclm} adopts a two-stage architecture that first predicts semantic tokens before rendering acoustics. 
MeLoDy~\cite{melody} enhances this design by replacing the acoustic decoder with a diffusion model to further improve synthesis quality and sampling efficiency.
In parallel, non-autoregressive methods~\cite{musicLDM, stableaudio, stableaudioopen} have emerged as promising alternatives. These models typically employ diffusion~\cite{DDPM, LDM} or flow-matching~\cite{flow-matching, I-CFM} mechanisms, operating in latent audio spaces to accelerate inference while maintaining high fidelity. 
Music ControlNet~\cite{music_control_net} extends this paradigm by introducing time-varying control signals—such as melody, rhythm, and dynamics—via temporally aligned conditioning, enabling fine-grained control over different aspects of the musical output. 
 TVC-MusicGen.~\cite{yang2025tvc} adopts a  segment-level diffusion framework for fine-grained music generation, sharing similarities with our paradigm, but its application is limited to instrumental music without vocal track.

\subsection{Song Generation}
Song generation aims to synthesize full musical compositions, including vocals and accompaniment, based on input lyrics and optional control signals.
This task introduces additional challenges beyond instrumental  music generation, such as aligning melody with lyrical phrasing, keeping vocal–accompaniment coherence, and maintaining musically meaningful transitions across segments.

Early systems such as Jukebox~\cite{jukebox} adopt autoregressive Transformers to model long-range dependencies over quantized audio tokens. Later works, like SongCreator~\cite{songcreator} and MusiCoT~\cite{music_cot},  follow similar strategies but rely on a shared token vocabulary for both vocal and instrumental content, which introduces modality interference and limits expressive capacity. 
To mitigate this, more recent systems, including SongGen~\cite{liu2025songgen}, YuE~\cite{yue} and LeVo~\cite{lei2025levo}, model vocals and accompaniment as separate token sequences, enabling more faithful synthesis and better exploitation of language models for long-context generation. However, these autoregressive approaches suffer from slow inference, high training complexity, and limited flexibility in downstream tasks such as song editing.
Alternatively, non-autoregressive frameworks, including DiffRhythm, DiffRhythm$+$, ACE-Step, and JAM~\cite{diffrhythm, diffrhythm_plus, acestepstep, liu2025jamtinyflowbasedsong}, leverage diffusion-based architectures for faster generation and easier extension. Nevertheless, these systems still face challenges to maintain musicality, long-range coherence, and balancing vocals with accompaniment, since both composition and acoustic rendering are handled within only  diffusion transformers.

In particular, existing systems rely primarily on global text prompts, which are insufficient to capture the temporal variability of real songs.  
Some works~\cite{yue, acestepstep, lei2025levo} insert  structural segmental labels in lyrics to align with predefined musical segments, yet finer-grained local control, such as adjusting the instrumentation or emotional intensity of specific segments, remains unsupported.
In contrast, our work introduces fine-grained, segment-level textual conditioning along with an LLM-based duration prediction module. 
These designs enable precise control over the temporal evolution of musical attributes while supporting  textual prompts for each segment, paving the way for more expressive and controllable song generation.

\FloatBarrier
\section{METHODOLOGY}
\begin{figure}[!t]
    \centering
    \includegraphics[width=\linewidth]{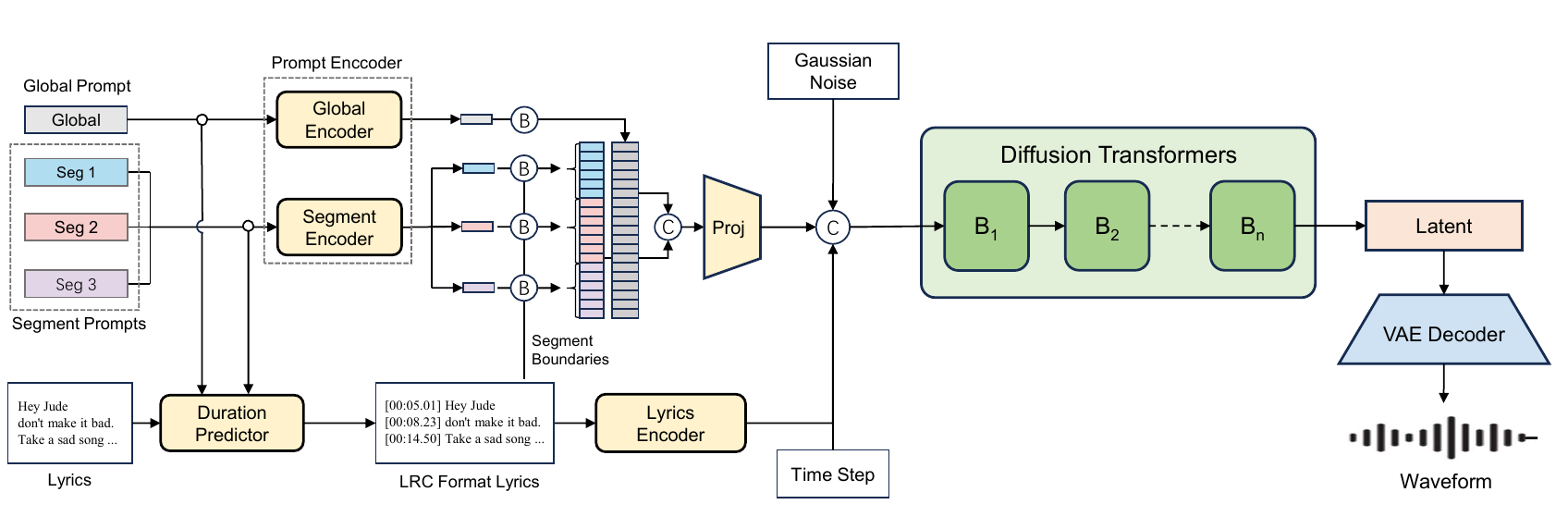}
    \caption{Overview of the \modelname{} architecture. The model takes lyrics and textual prompts as input. 
    A LLM-based duration predictor estimates sentence-level durations for the lyrics, while a lyrics encoder embeds the lyrics and performs sentence-level  alignment. 
    The prompt encoder encodes both global and segment prompts into 1D vectors. The global prompt is broadcast to all time steps, whereas each segment prompt is broadcast to the frames within its corresponding temporal window, as determined by the duration predictor.
    All the conditional embeddings are concatenated and fed into a Diffusion Transformer. In the diagram, \protect\circled{B} denotes temporal broadcasting within a segment window, and \protect\circled{C} denotes channel-wise feature concatenation.}
    \vspace{-1em}
    \label{fig:main}
\end{figure}

\subsection{Model Architecture}
\modelname{} adopts a Diffusion Transformer~\cite{DIT} architecture based on conditional flow-matching, extending previous work in flow-based music generation \cite{acestepstep,diffrhythm,diffrhythm_plus,liu2025jamtinyflowbasedsong}. The backbone consists of 16 LLaMA-style~\cite{llama} Transformer blocks.
The flow-matching process is conducted in the latent space. 
During training, a 1D VAE~\cite{stableaudio} is used to compress raw audio with a sampling rate of 44\,kHz into a latent sequence at 21.5\,Hz, which serves as the target trajectory for flow supervision.
Lyrics and textual descriptions are used as conditional inputs to control both the content and the stylistic characteristics of the generated music. 

In contrast to prior approaches, \modelname{} supports not only global textual conditioning but also segment-level control. 
This fine-grained conditioning is essential for improving both controllability and musical coherence.
In the following sections, we first describe the mechanism for segment-level textual control. 
We then introduce the sentence-level duration prediction module, which produces precise timestamp predictions at the sentence level, thereby addressing the limitations of previous DiT-based methods that relied on manually specified timestamps.

\subsubsection{Adding Segment-level Text Controls to Model Conditioning}
\modelname{} supports text conditioning at both the global and segment levels. 
The global prompt controls high-level song attributes such as genre, vocal gender and timbre, and the overall emotional tone. 
In contrast, segment-level prompts control time-varying attributes, including structural segment labels, local emotional change, and instrumentation.
During inference, segment prompts can be either specified by users or automatically generated by LLM in the prompt engineering stage.
Both global and segment prompts are encoded by the prompt encoder into a unified textual embedding $E_{\text{text}}$ with Algorithm~\ref{alg:text-conditioning}.
Specifically, a segment prompt is encoded by the segment encoder into a vector $\mathbf{e}_l^{i} \in \mathbb{R}^{1 \times d_l}$. 
The segment vector is then broadcast across segment’s temporal window, ensuring that each frame within the segment receives the local control information. 
Meanwhile, the global prompt is encoded by a global encoder and broadcast across all frames.
The global feature $E_g$ and segment feature $E_l$ are concatenated along the channel dimension and passed through a three-layer MLP projection module to produce the final conditioning embedding $E_{\text{text}} \in \mathbb{R}^{T \times d_{\text{text}}}$, where $T$ denotes the length of the latent sequence and $d_{\text{text}}$ is set to 1024.
We use  Qwen3-Embedding-0.6B~\cite{qwen3embedding} as the global and local prompt encoder, which is good at preserving fine-grained sementic attributes in the long document, and therefore suitable for our task.

Regarding lyric conditioning, \modelname{} first applies the duration predictor (described in the next section) to convert the raw lyrics into an LRC-style format with sentence-level timestamps.
The LRC-formatted lyrics are then fed into the lyrics encoder, which follows the sentence-level alignment strategy  in \cite{diffrhythm, diffrhythm_plus}. 
This alignment strategy improves rhythmic and prosodic control during generation while reducing the word error rate~\cite{liu2025jamtinyflowbasedsong}. 
The final output of the lyrics encoder    $E_{\text{lyrics}} \in \mathbb{R}^{T \times d_{\text{lyrics}}}$ is a sequence of length $T$.
Finally, the text prompt embedding ($E_{\text{text}}$),  lyrics embedding ($E_{\text{lyrics}}$), the audio latent sequence ($E_{\text{audio}}$), and the  time-step embedding $E_t$ (broadcast to match sequence length) are concatenated along the channel dimension and fed into the Diffusion Transformer as the input feature.

\begin{algorithm}[t]
\caption{Global and Segment Text Conditioning Embedding Extraction}
\setstretch{1.2}
\label{alg:text-conditioning}
\begin{algorithmic}[1]
\REQUIRE Global prompt $x_g$, segment prompts $\{(t_{s}^{i}, t_e^{i}, x_l^{i})\}_{i=1}^N$ 
\REQUIRE Sampling rate $r$, downsample rate $r_d$, number of latent frames $T$
\REQUIRE Global encoder $f_g$, local encoder $f_l$, output projection $\texttt{out\_proj}$

\STATE $\mathbf{e}_g \leftarrow f_g(x_g) \in \mathbb{R}^{1 \times d_g}$
\STATE $E_g \leftarrow \texttt{repeat}(\mathbf{e}_g, T)  \in \mathbb{R}^{T \times d_g} $   \hfill // Broadcast the global prompt across all time frames.  

\STATE Initialize $E_l \in \mathbb{R}^{T \times d_l}$ with zeros

\FOR{each $(t_{s}^{i}, t_e^{i}, x_l^{i})$ in segment prompts}
    \STATE $j_s^{i} \leftarrow \lfloor t_s^{i} \cdot r / r_d \rfloor$, $i_e \leftarrow \lfloor t_e \cdot r / r_d \rfloor$
    \STATE $\mathbf{e}_l^{i} \leftarrow f_l(x_l^{i}) \in \mathbb{R}^{1 \times d_l}$
    \STATE $E_l[j_s^{i} : j_e^{i}] \leftarrow \mathbf{e}_l^{i}$ \hfill // Broadcast each segment-level prompt to its corresponding temporal window. 
\ENDFOR

\STATE $E_{\text{cat}} \leftarrow \texttt{concat}(E_g, E_l, \text{dim}=-1) \in \mathbb{R}^{T \times (d_g + d_l)}  $
\STATE $E_{text} \leftarrow \texttt{out\_proj}(E_{\text{cat}})$
\RETURN fused embedding $E_{text} \in \mathbb{R}^{T \times d_{text}}$
\end{algorithmic}
\end{algorithm}

\subsubsection{Sentence-Level Duration Prediction}
\label{sec:dur_predictor}
As described above, duration prediction plays a critical role in the condition encoding stage of \modelname{}.
\begin{itemize}
    \item During inference, the total duration of the song is required to determine initial noise's length to be sampled;
    \item For lyric encoding, sentence-level timestamps are required to align the lyrics with the audio latent;
    \item For conditional encoding, segment durations are required to  enable window-based broadcasting.
\end{itemize}

Despite its importance, algorithm of duration prediction has been largely underexplored in previous NAR-based song generation works. 
Some works \cite{acestepstep,diffrhythm,diffrhythm_plus} require users to manually provide the total song duration or even sentence level timestamps.
This is not only inconvenient but also often leads to suboptimal generation quality, since user-provided timestamps are rarely good enough, especially when the user is not a professional musician.
JAM~\cite{liu2025jamtinyflowbasedsong} employs an in-context learning strategy to obtain word-level timestamps from LLMs in a zero-shot manner.
However, the zero-shot approach depends on complex prompt engineering, and the resulting predictions remain limited in quality, as accurate time prediction requires an inherent understanding of music and lyrics.

In this work, we fine-tune a LLM to act as a “composer” specifically for generating sentence-level lyric timestamps. We use Qwen3-4B-Base~\cite{qwen3technicalreport} as the base model, which strikes a favorable balance between model capacity and inference speed for this structured prediction task. 
The instruction template is provided in the Appendix \ref{appendix:B}. The input of duration predictor includes the lyrics as well as the global and local textual prompt, while the output is an LRC-style lyric file in which each line of lyrics is preceded by a sentence-level timestamp.
During training, the model learns to predict lyrical duration conditioned on the lyrical content and the musical attributes, including rhythm, emotion, and genre, which is crucial for accuratly duration predictions.

Note that the temporal windows for all segments can be naturally obtained from the sentence-level timestamps. For segments containing lyrics (e.g., verse or chorus), the corresponding temporal windows are directly derived from the sentence-level timestamps. For instrumental segments without lyrics (e.g., intro, bridge, or outro), their temporal boundaries are inferred from the adjacent lyric-containing segments, since such regions are always contiguous in song structure.

Experimental results show that our duration predictor yields more precise sentence-level timing and achieves higher musical aesthetic scores compared with zero-shot LLM timestamp prediction, which proves the effectiveness of the proposed method.

\subsection{Data Pipeline}
\subsubsection{Quality Filtering}

The raw corpus varies widely in both recording fidelity and musical quality.  
To ensure data quality, we employ a two-stage quality filtering process that combines simple signal processing with deep-learning-based strategy.  
In the beginning of metadata filtering stage, the sound evnet detection module is employed to remove non-musical samples.  
Records are then filtered by acoustic metadata including duration, sampling rate, channel numbers, compression rate and energy.

Following metadata filtering, we conduct automatic quality assessment using two open-source evaluation frameworks: Audiobox Aesthetics~\cite{audiobox_Aesthetics} and SongEval~\cite{songeval}.  
The former provides general audio aesthetic scores, while the latter is originally designed for generative music quality assessment, but we find that it is also effective for real-song aesthetics evaluation.

\subsubsection{Lyrics Processing}

Lyric transcription is necessary for songs that lack lyric annotations.  
In the pipeline, We first apply Demucs v4~\cite{demucs} for vocal–accompaniment separation, followed by automatic speech recognition (ASR) on the vocal track.  
FireRedASR~\cite{fireredasr} is used for Mandarin, while other languages are transcribed by Whisper-Large-v3~\cite{whisper}.
Some records in the corpus include pre-existing LRC-formatted lyric files.  
For these records, an LLM-based cleaner is first employed to remove song metadata in the lrc file.  
The cleaned lyrics are then computed edit distance with ASR transcripts, and samples with high edit distance are discarded.  In addition to lyrics, we also need to extract structural segmentation labels (e.g., \textit{intro}, \textit{verse}, \textit{chorus}, \textit{outro}), which were identified using an all-in-one music understanding model~\cite{taejun2023allinone}.

\subsubsection{Music Caption}

Textual descriptions of songs are generated using Audio Flamingo 3~\cite{audioflamingo3}.  
For global captioning, the full audio are used as the input, while for segment-level captioning, each segment is captioned independently.  
Prompt templates used during inference are detailed in the Appendix \ref{appendix:B}.  
To implemnet segment controllability, the structural segmentation label from all-in-one model is prepended to the caption.  
Additionally, the first and last 0.5\,seconds of each song are assigned fixed prompt prompts “This piece is the start/end of the song.” to mark song boundaries.

\begin{figure}[!t]
    \centering
    \includegraphics[width=\linewidth]{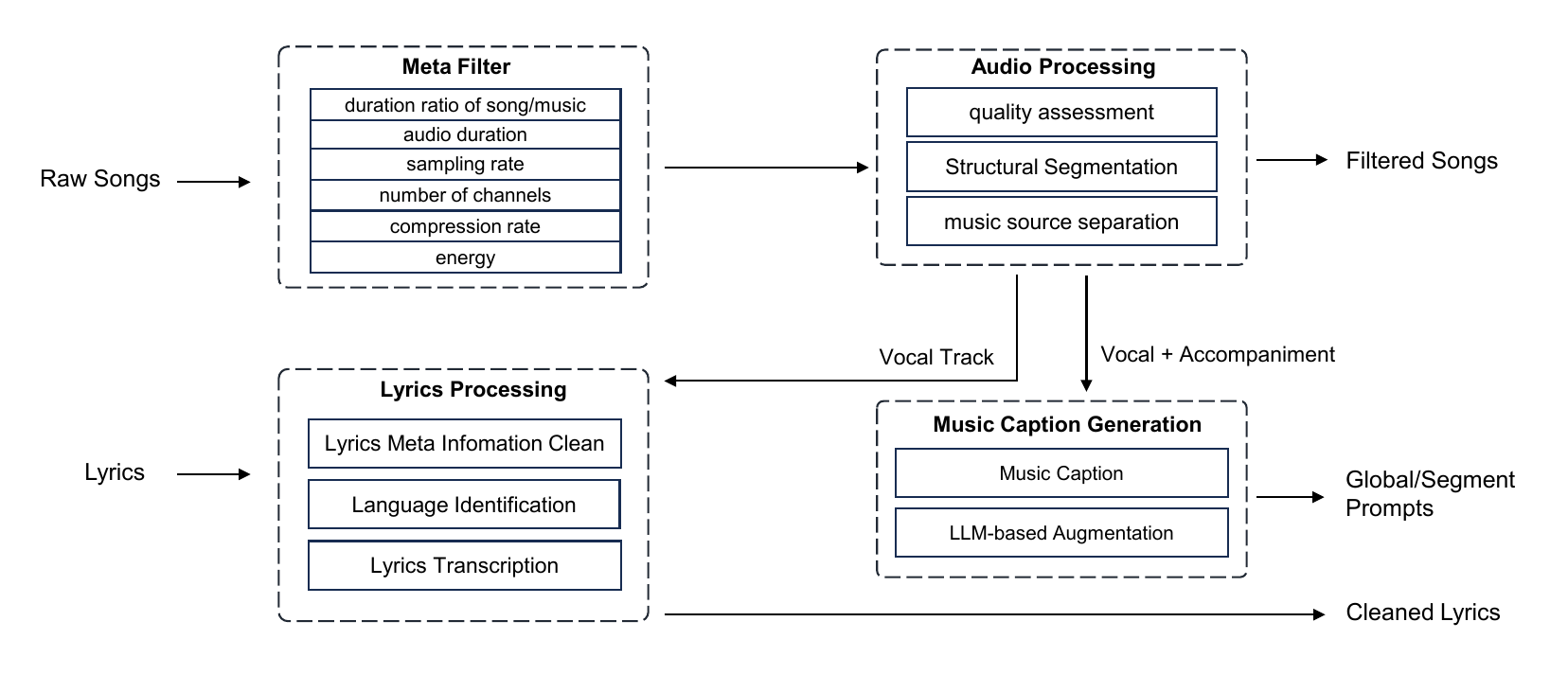}
    \vspace{-0.8em}
    \caption{Overview of the data pipeline of \modelname{}}
    \label{fig:data-pipeline}
\end{figure}

\modelname{} is trained on a curated in-house corpus primarily composed of Chinese pop songs.  
We established a structured data pipeline to clean, organize, and annotate raw musical content.
An overview of the pipeline is illustrated in Figure~\ref{fig:data-pipeline}.

\subsection{Training and Inference}

The model is trained under the Conditional Flow Matching (CFM) framework, which aims to learn a function $v_\theta(t, C, x_t)$ that approximates the flow $u(x_t|x_0,x_1)$.  
The training objective is defined as:
\begin{align}
&\mathcal{L}_{\text{CFM}}(\theta) = 
\mathbb{E}_{t,\, q(x_1),\, p(x_0)} 
\left\| v_\theta(t, C, x_t) - u(x_t \mid x_0, x_1) \right\|^2 \\
&x_t = (1 - t)x_0 + t x_1 \\
&u(x_t \mid x_0, x_1) = x_1 - x_0
\end{align}
Here, $x_0 \sim \mathcal{N}(0, \mathbf{I})$ represents a sample from the prior distribution,  
$x_1 \sim q(x_1)$ is drawn from the target data distribution,  
$t \sim \mathcal{U}(0, 1)$ denotes the diffusion time step,  
and $C$ denotes the conditioning input, which includes lyrics and textual prompts.  
The target vector $u(x_t \mid x_0, x_1)$ represents the flow at $x_t$.

The training procedure follows a three-stage paradigm:

\begin{enumerate}
    \item \textbf{Pre-training:}  
    In this stage, we filter out songs with sampling rates lower than 32\,kHz and restrict song durations to the range from 30\,s to 6\,min.  
    Additionally, we remove the lowest 5\% of samples ranked by the automatic quality assessment score.  
    The resulting pre-training dataset contains approximately 370{,}000 songs (around 27{,}000 hours of audio).

    \item \textbf{Fine-tuning:}  
    In the fine-tuning stage, stricter filtering criteria are applied.  
    Songs are required to have a 44\,kHz sampling rate, stereo channels, and belong to the top 50\% according to all automatic quality assessment metrics in the corpus.  
    After filtering, around 50{,}000 songs (roughly 4{,}000 hours) are used for fine-tuning.

    \item \textbf{Preference Alignment:}  
    As demonstrated in previous works~\cite{lei2025levo,diffrhythm_plus, liu2025jamtinyflowbasedsong},  
    Direct Preference Optimization~\cite{dpo} (DPO) is effective in improving the perceptual quality of music generation.  
    We adopt the iterative DPO strategy following~\cite{liu2025jamtinyflowbasedsong}. 
    The SFT model is set as the initial policy and then three iterative rounds of DPO are conducted.
    In each round, we first use the model weight from the previous round to generate 16 songs for each lyric sample.  
    Win–loss pairs are constructed by selecting samples whose SongEval\cite{songeval} score difference exceeds a threshold and whose win sample’s SongEval score lies above the third quartile of all the sampled songs. 
    Approximately 20,000 win–loss pairs are included in each round of DPO training.
\end{enumerate}

During inference, we employ the Euler ODE solver.  
For Classifier-Free Guidance (CFG), we use the following formulation:
\begin{align}
v = v_u + \text{cfg}(v_c - v_u) - \text{cfg}_n(v_n - v_u)
\end{align}
where $v_u$ and $v_c$ denote the unconditional and conditional flows, respectively,  
and $v_n$ represents the flow obtained under negative conditions~\cite{negative_prompt}.  
In the negative condition setup, the lyric conditioning is removed,  
while both global and local prompts are replaced with negative prompts.  
Empirically, the coefficients are set to $\text{cfg}=3$ and $\text{cfg}_n=1$.

\FloatBarrier
\section{EXPERIMENTALS}

\subsection{Experimental Setup}



\modelname{} is trained on an in-house dataset comprising over 90\% Mandarin pop music, covering a wide range of artist identities, lyrical themes, and musical structures. The diffusion backbone adopts a DiT-style architecture with approximately 1.1B parameters and 16 LLaMA-style decoder blocks, following~\cite{diffrhythm}.

The training pipeline consists of three stages. We first pretrain the model for 20 epochs using a batch size of 32 and a learning rate of 2e-5, followed by fine-tuning under the same settings for 8 epochs. 
Then in the 3-rounds iterative DPO stage, we train the model for 4 epochs with batch size 8, gradient accumulation of 4, and a learning rate of 5e-7 in each round.
During training, 20\% dropout is independently applied to both global and segment-level conditions to support classifier-free guidance. 
We also augment global and segment text prompts using LLM-based rewriting to enhance generalization across diverse real-world text input styles.

For the duration prediction module, we fine-tune Qwen3-4B-Base  using approximately 100,000 LRC-format lyrics.
Training is conducted for 8 epochs with a batch size of 8, gradient accumulation of 4, max new tokens of 4096 and a learning rate of 2e-5. 
The LoRA~\cite{hu2022lora} strategy with a rank of 32 is applied to improve training efficiency.

\subsection{Evaluation Metrics}

The test set is constructed using 15 Mandarin pop lyrics generated by ChatGPT. 
To comprehensively assess the quality of generated content, we employ Audiobox-aesthetic~\cite{audiobox_Aesthetics} and SongEval~\cite{songeval} as content-oriented evaluation metrics. Audiobox-aesthetic examines four dimensions: production quality (PQ)—which considers clarity, fidelity, dynamics, and spatialization; production complexity (PC)—reflecting the richness of audio components; content enjoyment (CE)—capturing subjective appeal and artistic expressiveness; and content usefulness (CU)—indicating the potential for reuse in creative workflows.
In addition, SongEval provides a more holistic assessment of musical generation by measuring coherence (Coh), memorability (Mem), naturalness of vocal breathing and phrasing (NVBP), clarity of song structure (CSS), and overall musicality (OM). Together, these metrics capture both the technical and aesthetic dimensions of the generated music.

The instruction-following capability of music generation models is then evaluated using the MuLan~\cite{zhu2025muq} score. Muq-Mulan is a widely adopted multimodal representation model that aligns text descriptions and music clips through contrastive learning, and has been extensively used for music evaluation \cite{acestepstep,yue}. Global Mulan score measures the global alignment between generated songs and their corresponding textual prompts. In addition, we compute the Segment MuLan scores for individual song segments based on segment prompts and use their average to assess the model’s segment instruction adherence.

Beyond MuLan score, we further extend our evaluation to typical singer-related musical attributes, such as gender and age, since we find that Mulan score fails to reflect singer characteristics.
Prompts are selected from the test set, and the singer’s gender and age (teenager, 20s, and 40s) are modified to form new prompts. For gender evaluation, we directly employ Qwen3-Omni-30B-A3B-Captioner~\cite{Qwen3-Omni} to assess the gender of generated songs. For age evaluation, we randomly sample two generated songs and conduct an A/B test, where Qwen3-Omni determines which song corresponds to the older singer. The final accuracy is computed based on the correctness of these pairwise judgments.

\section{RESULTS}

\subsection{Results of Objective Evaluation}



To compare the performance of SegTune against recent open-source song generation systems, we selected four representative, state-of-the-art open-source models: YuE~\cite{yue}, LeVo~\cite{lei2025levo}, DiffRhythm$+$~\cite{diffrhythm_plus}, and ACE-Step~\cite{acestepstep}. Among these, YuE and LeVo employ language models as their primary generation architecture, whereas DiffRhythm$+$ and ACE-Step are representative of diffusion-based architectures. All these baseline models support both global textual tags and lyrics as control conditions for song generation.

\paragraph{Music Aesthetic Evaluation}
Table~\ref{tab:performance_comparison} reports the objective evaluation results of the SegTune model at both the SFT and DPO stages, alongside several baseline models, across key dimensions of musicality, complexity, and naturalness, as measured by AudioBox-aesthetic and SongEval.
As shown in Table~\ref{tab:performance_comparison}, SegTune demonstrates strong performance across both AudioBox-aesthetic and SongEval metrics. SegTune-SFT achieves competitive results compared to prior methods, while SegTune-DPO further improves musical quality, particularly in coherence and structure-related dimensions.

It is worth noting that SegTune is trained on a Mandarin pop-centric dataset, while the baselines are designed as general-purpose music generation models. To further isolate the contribution of our proposed techniques, we conduct controlled ablation studies in later sections under comparable experimental settings.

\paragraph{Instruction-following} Given that baseline models only support global tag control, we report Global MuLan scores, gender and age control accuracy to assess instruction adherence (Table~\ref{tab:struction_follow_baseline}).
SegTune-SFT achieves the highest gender control accuracy among all models while maintaining strong MuLan alignment, indicating its superior ability to follow textual instructions.
In comparison, SegTune-DPO preserves high musical quality but exhibits reduced gender control accuracy, likely due to the preference pair selection process overlooking instruction-following constraints and SongEval’s implicit bias toward female vocals.

\newcolumntype{Y}{>{\centering\arraybackslash}X}
\begin{table}[htbp]
\centering
\small
\caption{Performance comparison of SegTune and baseline models on AudioBox-aesthetic and SongEval metrics. SegTune-SFT denotes the model that has undergone both pretraining and supervised fine-tuning (SFT), while SegTune-DPO refers to the model further refined via 2 iterations of Direct Preference Optimization (DPO) starting from the SegTune-SFT checkpoint.}
\label{tab:performance_comparison}
%
\begin{tabularx}{0.93\textwidth}{@{} >{\centering\arraybackslash}p{1.5cm} *{9}{>{\centering\arraybackslash}p{0.65cm}} @{}}
\toprule
\addlinespace[4pt] 
\multirow{4}{*}{\textbf{Models}}  & \multicolumn{4}{c}{\textbf{AudioBox-aesthetic}} & \multicolumn{5}{c}{\textbf{SongEval}} \\
\addlinespace[4pt] 
\cmidrule(lr){2-5} \cmidrule(lr){6-10}
\addlinespace[4pt] 
&  \multicolumn{1}{c}{\textbf{CE$\uparrow$}} & \multicolumn{1}{c}{\textbf{CU$\uparrow$}} & \multicolumn{1}{c}{\textbf{PC$\uparrow$}} & \multicolumn{1}{c}{\textbf{PQ$\uparrow$}} & \multicolumn{1}{c}{\textbf{Coh$\uparrow$}} & \multicolumn{1}{c}{\textbf{Mem$\uparrow$}} & \multicolumn{1}{c}{\textbf{NVBP$\uparrow$}} & \multicolumn{1}{c}{\textbf{CSS$\uparrow$}} & \multicolumn{1}{c}{\textbf{OM$\uparrow$}} \\
\addlinespace[4pt] 
\midrule
\addlinespace[6pt] 
YuE  & 7.16 & 7.66 & 6.27 & 8.09 &3.51  & 3.27 & 3.22 & 3.26 & 3.22\\
\addlinespace[4pt] 
LeVo &  7.43 & 7.71 & 5.25 & \underline{8.29} & 3.46 & 3.29 & 3.20 & 3.29 & 3.35 \\
\addlinespace[4pt] 
DiffR.$^+$ &  \underline{7.55} & \underline{7.80} & 6.72 & 8.21& \underline{4.05} & \underline{3.84} & \underline{3.65} & \underline{3.82} & \underline{3.76}   \\
\addlinespace[4pt] 
ACE-Step &  7.38 & 7.53 & 6.71 & 7.88 & 3.98 & 3.78 & \underline{3.65} & 3.77 & 3.74 \\
\addlinespace[4pt] 
 \hdashline
\addlinespace[6pt] 
SegTune &    &  &  &  &  &  &  &  &  \\[2ex]
\quad  - SFT &  7.38  & 7.71 & \textbf{6.83} & 8.23 & 3.54 & 3.22 & 3.23 & 3.32 & 3.19\\
\addlinespace[4pt]
\quad  - DPO &   \textbf{7.63} & \textbf{7.85} & \underline{6.80} & \textbf{8.36} & \textbf{4.25} & \textbf{4.06} & \textbf{4.09} & \textbf{4.08} & \textbf{3.97}\\

\addlinespace[4pt] 
\bottomrule
\end{tabularx}
\end{table}

\begin{table}[htbp]
\centering
\vspace{1.5em}
\caption{Performance of SegTune and baseline models on instruction-following metrics. }
\label{tab:struction_follow_baseline}
\begin{tabular}{lcccccc}
\toprule
Metrics & YuE & LeVo  & DiffR$^+$ & ACE-Step  & SegTune-SFT & SegTune-DPO \\
\midrule
Global Mulan$\uparrow$ & 0.29  & 0.32 & \textbf{0.47} & 0.35  & \textbf{0.47} & \underline{0.46} \\
Gender (\%)$\uparrow$ & 80.65 & \underline{90.60} & 37.50 & 78.12  &   \textbf{96.67} & 80.95 \\ 
Age (\%)$\uparrow$ & 44.00 & 50.00 & 54.00 & \underline{56.00} & \textbf{57.00} &  51.00    \\ 
\bottomrule
\end{tabular}
\end{table}

\subsection{Ablation Studies}
\subsubsection{Ablation for Prompt Encoder}

Since the SegTune model differs from other baselines in terms of training data, parameter scale, and model architecture, the strong performance of SegTune-SFT and SegTune-DPO reported in Table~\ref{tab:performance_comparison} alone cannot conclusively demonstrate the superiority of the segment control strategy over global control. To enable a more rigorous and fair evaluation of the effectiveness of segment-level prompt injection, we conduct ablation studies using identical training data and compare models equipped with different prompt encoder settings under the same experimental setup. 

In the ablation studies, we not only investigate various composition strategies of prompt encoders but also compare the impact of two encoder choices—Qwen3-Embedding and Muq-Mulan—on music generation.
Qwen3-Embedding represents the current state-of-the-art in text embedding and is particularly well-suited to SegTune, where prompts consist of natural-language, long-form textual descriptions. 
In contrast, as previously introduced, Muq-Mulan is a multimodal representation model specifically trained on music data and has already been widely adopted in music generation \cite{ diffrhythm_plus, liu2025jamtinyflowbasedsong, yang2025tvc}.

\newcolumntype{C}{>{\centering\arraybackslash}p{0.5cm}}
\begin{table}[t]
\centering
\small
\caption{Impact of various prompt encoder settings on objective performance of SegTune at the SFT stage. MuQ refers to MuQ-Mulan encoder, and Qwen3. refers to Qwen3-Emebedding.}
\label{tab:performance_comparison_ablation}

\begin{tabularx}{\textwidth}{>{\centering\arraybackslash}p{0.9cm} >{\centering\arraybackslash}p{0.9cm} *{9}{Y} @{}}
\toprule
\addlinespace[6pt] 
\multicolumn{2}{c}{\textbf{Prompt encoders}}& \multicolumn{4}{c}{\textbf{AudioBox-aesthetic}} & \multicolumn{5}{c}{\textbf{SongEval}} \\
\addlinespace[4pt] 
\cmidrule(lr){1-2} \cmidrule(lr){3-6} \cmidrule(lr){7-11}
\addlinespace[4pt] 
\multicolumn{1}{c}{\textbf{Global}}& \multicolumn{1}{c}{\textbf{Segment}} & \textbf{CE$\uparrow$} & \textbf{CU$\uparrow$} & \textbf{PC$\uparrow$} & \textbf{PQ$\uparrow$} & \multicolumn{1}{c}{\textbf{Coh$\uparrow$}} & \multicolumn{1}{c}{\textbf{Mem$\uparrow$}} & \multicolumn{1}{c}{\textbf{NVBP$\uparrow$}} & \multicolumn{1}{c}{\textbf{CSS$\uparrow$}} & \multicolumn{1}{c}{\textbf{OM$\uparrow$}} \\
\addlinespace[4pt] 
\midrule
\addlinespace[4pt] 
 \multicolumn{2}{l}{\textbf{Global-only}}& & & & & & & & &  \\[2ex]
 \multicolumn{1}{c}{MuQ} & \multicolumn{1}{c}{--}      & \underline{7.42} & 7.60 & 6.63 & 8.19 & 3.18 & 2.87 & 2.81 & 2.93 & 2.86 \\
 \addlinespace[4pt] 
 \multicolumn{1}{c}{Qwen3.} & \multicolumn{1}{c}{-- }   & 7.39 & 7.64 & \underline{6.76} & 8.19 & 3.42 & 3.11 & 3.11 & 3.20 & 3.12 \\
 \addlinespace[4pt] 
 \hdashline
 \addlinespace[4pt] 
 \multicolumn{2}{l}{\textbf{Concat.}}& & & & & & & & &  \\[2ex]
 \multicolumn{1}{c}{Qwen3.} & \multicolumn{1}{c}{MuQ}   & \textbf{7.57} & \textbf{7.82} & 6.63 & \textbf{8.35} & \textbf{3.62} & \textbf{3.37} & \textbf{3.30} & \textbf{3.43} & \textbf{3.34} \\
 \addlinespace[4pt] 
 \multicolumn{1}{c}{Qwen3.} & \multicolumn{1}{c}{Qwen3.} & 7.38  & 7.71 & \textbf{6.83} & 8.23 & \underline{3.54} & \underline{3.22} & \underline{3.23} & \underline{3.32} & \underline{3.19} \\
 \addlinespace[4pt] 
 \hdashline
 \addlinespace[4pt] 
 \multicolumn{2}{l}{\textbf{Mixed}}& & & & & & & & &  \\[2ex]
 \multicolumn{1}{c}{Qwen3.} & \multicolumn{1}{c}{Qwen3.} & 7.29 & \underline{7.73} & 6.33 & \underline{8.32} & 3.43 & 3.14 &3.15  &3.23  &3.12  \\
\addlinespace[4pt] 
\bottomrule
\end{tabularx}
\end{table}

\begin{table}[t]
\centering
\small
\vspace{1.3em}
\caption{Impact of various prompt encoder settings on instruction-following performance of SegTune at the SFT stage. MuQ refers to MuQ-Mulan encoder, and Qwen3. refers to Qwen3-Emebedding.}
\label{tab:instruct_follow_ablation}

\begin{tabularx}{0.75\textwidth}{>{\centering\arraybackslash}p{0.9cm} >{\centering\arraybackslash}p{0.9cm} *{4}{>{\centering\arraybackslash}p{1.5cm}} @{}}
\toprule
\addlinespace[6pt] 
\multicolumn{2}{c}{\textbf{Prompt encoders}}& \multicolumn{4}{c}{\textbf{Instruction-following}} \\
\addlinespace[4pt] 
\cmidrule(lr){1-2} \cmidrule(lr){3-6} 
\addlinespace[4pt] 
\multicolumn{1}{c}{\textbf{Global}}& \multicolumn{1}{c}{\textbf{Segment}} & \textbf{Global Mulan$\uparrow$} & \textbf{Segment Mulan$\uparrow$} & \textbf{Gender$\uparrow$} & \textbf{Age$\uparrow$}  \\
\addlinespace[2pt] 
\midrule
\addlinespace[4pt] 
 \multicolumn{2}{l}{\textbf{Global-only}}& & & &  \\
 \addlinespace[4pt] 
 \multicolumn{1}{c}{MuQ} & \multicolumn{1}{c}{--}      & 0.389 & 0.300 & 47.6\% & 47\%  \\
 \addlinespace[4pt] 
 \multicolumn{1}{c}{Qwen3.} & \multicolumn{1}{c}{--}    & 0.401 & 0.328 & \underline{92.2\%} & 50\%  \\
 \addlinespace[4pt] 
 \hdashline
 \addlinespace[4pt] 
 \multicolumn{2}{l}{\textbf{Concat.}}& & & &    \\
 \addlinespace[4pt] 
 \multicolumn{1}{c}{Qwen3.} & \multicolumn{1}{c}{MuQ}   & \underline{0.436} & \underline{0.373} & 84.4\% & 46\% \\
 \addlinespace[4pt] 
 \multicolumn{1}{c}{Qwen3.} & \multicolumn{1}{c}{Qwen3.} & \textbf{0.465} & \textbf{0.378} & \textbf{96.7\%} & \textbf{57\%}  \\
 \addlinespace[4pt] 
 \hdashline
 \addlinespace[4pt] 
 \multicolumn{2}{l}{\textbf{Mixed}}& & & &  \\
 \addlinespace[4pt] 
 \multicolumn{1}{c}{Qwen3.} & \multicolumn{1}{c}{Qwen3.} & 0.434 & 0.346 & 90.5\% & \textbf{57\%}  \\
\addlinespace[4pt] 
\bottomrule
\end{tabularx}
\end{table}

\paragraph{Global-only Setting} First, we retain only the global prompt and its corresponding encoder for model training and conduct objective evaluation on the test set. As shown in Table~\ref{tab:performance_comparison_ablation}, the model using Qwen3-Embedding as global encoder consistently outperforms the one using the MuQ-Mulan~\cite{zhu2025muq} encoder in terms of general musicality. At the instruction-following level (see Table~\ref{tab:instruct_follow_ablation}), music generation controlled by the Qwen3-Embedding global encoder achieves higher scores on both the Global Mulan and Segment Mulan metrics. In particular, the Qwen3-Embedding-based global encoder improves the accuracy of the singer's gender control to 92.2\%, while the MuQ-Mulan global encoder exhibits almost no ability to control gender. This discrepancy stems primarily from the fact that during MuQ-Mulan's text–music alignment training, singer-related attributes were not included in music captions, resulting in a poor representation of vocal characteristics in its embeddings. 

\paragraph{Concatenate Setting} Given that Qwen3-Embedding, when used as the global encoder, demonstrates highly accurate control over singer gender, we fix it as the global encoder and subsequently compare the performance of two configurations for the segment prompt encoder: MuQ-Mulan versus Qwen3-Embedding. Although MuQ-Mulan exhibits weaker representations of singer-related attributes, it remains a viable candidate for the segment prompt encoder because segment-level prompts primarily govern musical attributes such as instrumentation, emotion, and rhythm—domains in which MuQ-Mulan is still effective.

As shown in Table~\ref{tab:performance_comparison_ablation} and Table~\ref{tab:instruct_follow_ablation}:
(1) The two concatenate settings achieve comparable overall musicality, yet the configuration employing Qwen3-Embedding as the segment prompt encoder yields superior instruction-following performance.
(2) Both objective musicality and the instruction follow-up capabilities of the concatenate settings are significantly higher than those of the global-only setting.
These results provide strong empirical evidence supporting the rationale for SegTune's architectural design.

\paragraph{Mixed Setting} Following the feature fusion framework of \cite{jiang2025freeaudio}, we also explored the linear mixture strategy to fuse global and segment embeddings, assigning a weight of 0.2 to the global prompt embedding and 0.8 to the segment prompt embedding. However, compared to the concatenate setting, this mixed approach leads to a noticeable degradation in both musicality and instruction-following accuracy. This decline is primarily attributed to the fact that the mixed setting  obscures the distinction between global prompts and segment prompts.

\subsubsection{Ablation for Duration Predictor}
As introduced in Section~\ref{sec:dur_predictor}, an accurate duration predictor can generate lyric timestamps that are more musically plausible than those provided by human users, thereby enhancing the overall quality of the generated music and reducing the need for user expertise. We evaluated two duration prediction approaches: (1) a Qwen3-4B-Base model, post-trained on our dataset, to predict the start timestamp of each lyric line; and (2) a zero-shot predictor using GPT-4o with a carefully designed prompt, following JAM~\cite{liu2025jamtinyflowbasedsong}. 
Unlike previous experiments that used synthetic lyrics, this evaluation uses 15 real-world Mandarin pop songs with paired lyrics and prompts, to enable direct comparison with ground-truth timestamps. These songs are excluded from the training set to ensure fair evaluation.

We compared the musicality and instruction-following accuracy of songs generated by SegTune-DPO when conditioned on either the predicted timestamps or the ground-truth timestamps, respectively. As shown in Table~\ref{tab:dur_predictor}, the Qwen3-SFT duration predictor achieves a mean absolute error (MAE) of 0.99 seconds, substantially lower than that of the zero-shot GPT-4o predictor. Moreover, while Qwen3-SFT performs comparably to GPT-4o on the Content Enjoyment and Content Usefulnes, it consistently outperforms GPT-4o across all other musicality metrics.

Table~\ref{tab:dur_predictor_instruction_following} shows that a higher MAE in timestamp prediction leads to lower global Mulan scores. This is mainly because inaccurate timings of lyric onset adversely affect the overall musical arrangement—for example, causing certain lyric lines to be rendered at unnaturally fast or slow singing rates. However, this degradation has a relatively minor impact on segment Mulan scores and does not compromise the accuracy of gender control.

\begin{table}[htbp]
\centering
\small
\caption{Impact of duration predictor on SegTune-DPO performance. MAE denotes the mean absolute error (in seconds) of sentence-level duration prediction. GT means using the ground truth timestamps of lyrics for model inference.}
\label{tab:dur_predictor}

\begin{tabularx}{\textwidth}{@{} >{\centering\arraybackslash}p{1.5cm} >{\centering\arraybackslash}p{0.8cm} *{9}{>{\centering\arraybackslash}p{0.55cm}} @{}}
\toprule
\addlinespace[4pt] 
\multirow{4}{*}{\textbf{Models}} & \multirow{4}{*}{\textbf{MAE}}& \multicolumn{4}{c}{\textbf{AudioBox-aesthetic}} & \multicolumn{5}{c}{\textbf{SongEval}} \\
\addlinespace[4pt] 
\cmidrule(lr){3-6} \cmidrule(lr){7-11}
\addlinespace[4pt] 
& &  \multicolumn{1}{c}{\textbf{CE$\uparrow$}} & \multicolumn{1}{c}{\textbf{CU$\uparrow$}} & \multicolumn{1}{c}{\textbf{PC$\uparrow$}} & \multicolumn{1}{c}{\textbf{PQ$\uparrow$}} & \multicolumn{1}{c}{\textbf{Coh$\uparrow$}} & \multicolumn{1}{c}{\textbf{Mem$\uparrow$}} & \multicolumn{1}{c}{\textbf{NVBP$\uparrow$}} & \multicolumn{1}{c}{\textbf{CSS$\uparrow$}} & \multicolumn{1}{c}{\textbf{OM$\uparrow$}} \\
\addlinespace[4pt] 
\midrule
\addlinespace[6pt] 
GT        & 0.00  & 7.65 & 7.74 & 6.58 & 8.35 & \textbf{4.33} & 4.16 & 4.17 & 4.12 & 4.01 \\
\addlinespace[4pt] 
Qwen3-sft & 0.99  & 7.66 & 7.74 & \textbf{6.58} & \textbf{8.36} & 4.32 & \textbf{4.16} & \textbf{4.18} & \textbf{4.16} & \textbf{4.06}\\
\addlinespace[4pt] 
GPT-4o    & 3.24  & \textbf{7.69} & \textbf{7.76} & 6.29 & 8.34 & 4.19 & 4.00 & 4.08 & 3.98 & 3.86\\
\addlinespace[4pt] 

\bottomrule
\end{tabularx}
\end{table}

\begin{table}[htbp]
\centering
\small
\caption{Impact of duration predictor on instruction-following performance of SegTune-DPO. MAE denotes the mean absolute error (in seconds) of sentence-level duration prediction. GT means using the ground truth timestamps of lyrics for model inference.}
\label{tab:dur_predictor_instruction_following}

\begin{tabularx}{0.77\textwidth}{>{\centering\arraybackslash}p{1.5cm} >{\centering\arraybackslash}p{1.0cm} *{4}{>{\centering\arraybackslash}p{1.5cm}} @{}}
\toprule
\addlinespace[6pt] 
\multirow{4}{*}{\textbf{Models}} & \multirow{4}{*}{\textbf{MAE}} & \multicolumn{4}{c}{\textbf{Instruction-following}} \\
\addlinespace[4pt] 
 \cmidrule(lr){3-6} 
\addlinespace[4pt] 
 & & \textbf{Global Mulan$\uparrow$} & \textbf{Segment Mulan$\uparrow$} & \textbf{Gender$\uparrow$} & \textbf{Age$\uparrow$}  \\
\addlinespace[4pt] 
\midrule
\addlinespace[4pt] 
 \multicolumn{1}{c}{GT} & 0.00& 0.453 & 0.466 &81.9\% & 61\%  \\
  \addlinespace[4pt] 
 \multicolumn{1}{c}{Qwen3-sft} & 0.99  & 0.453 & 0.410 & 81.9\% & 61\%  \\
 \addlinespace[4pt] 
 \multicolumn{1}{c}{GPT-4o} & 3.24  & 0.420 & 0.414 & 81.9\% & 58\%  \\
 \addlinespace[4pt] 
 
\bottomrule
\end{tabularx}
\end{table}


\FloatBarrier
\section{CONCLUSION}
In this work, we introduced \modelname{}, a non-autoregressive framework for structured and controllable song generation. 
By integrating segment-level textual conditioning and an LLM-based duration predictor, \modelname{} enables fine-grained control over musical attributes such as emotion, rhythm, and instrumentation, while maintaining global stylistic coherence. 
Together with a large-scale data pipeline and a unified evaluation protocol, our system achieves substantial improvements in controllability and musical aesthetics compared to existing  open-source baselines.

Looking forward, we plan to extend this framework along three directions. 
First, we aim to integrate our segment-level control mechanism into autoregressive architectures to leverage their strengths in long-sequnece modeling. 
Second, future work will focus on richer local control—such as enabling duet singing and multi-singer transitions—which remains challenging for \modelname{} due to data scarcity. 
Finally, we envision integrating SegTune into an interactive music-generation agent, where a conversational LLM can interpret user intent and dynamically compose lyrics and global-local prompts to guide the generation process.

\begin{ack}
We thank the following colleagues for their support and contributions to this work ~(sorted by alphabetical order): Youjun Chen, Feng Deng, Qianyue Hu, Nan Li, Zewen Song, Ming Wen, Junjie Yan, Kang Yin, Jingru Zhao.
\end{ack}

\bibliographystyle{unsrt}
\bibliography{references}

\newpage
\appendix

\section{Audio-Flamingo3 Caption Prompt} \label{appendix:A}
\begin{tcolorbox}[
  enhanced,
  colback=gray!5,
  colframe=black,
  coltitle=white,
  colbacktitle=black,
  title=\textbf{Global Caption Prompt},
  breakable,
  sharp corners,
  boxrule=0.5pt
]

You are a helpful AI assistant. You need to act as a caption generator for music and generate descripitons in MusicCaps style. Describe the music in vivid detail, using the following rules: 

\begin{enumerate}
    \item Describe the details about genre, mood, feeling, ambience, and other notable features of the music.
    \item Describe the singer's  vocal characteristics, including gender, age range, vocal timbre, pitch range, and other notable features of the singer.
    \item Keep the descripiton within 1-4 sentences.
    \item Only provide details you are confident about. It is not compulsory to provide all details, but do not hallucinate.
\end{enumerate}

\end{tcolorbox}

\begin{tcolorbox}[
  enhanced,
  colback=gray!5,
  colframe=black,
  coltitle=white,
  colbacktitle=black,
  title=\textbf{Segment Caption Prompt},
  breakable,
  sharp corners,
  boxrule=0.5pt
]

You are a helpful AI assistant. Describe the song segment as part of a complete piece of song in vivid detail according to what you hear. Generate the descripiton using the following rules: 

\begin{enumerate}
    \item Include the instrumentation, rhythm and melody style, mood, emotional's impact, intensity and change.
    \item Mention any notable singing and playing techniques that occur and dynamic changes of the song.
    \item Keep the descripiton within 1-3 sentences.
    \item Only provide details you are confident about. It is not compulsory to provide all details, but do not hallucinate.
    
\end{enumerate}

\end{tcolorbox}

\section{Prompt Example of Duration Predictor Module} \label{appendix:B}

\begin{tcolorbox}[
  enhanced,
  colback=gray!5,
  colframe=black,
  coltitle=white,
  colbacktitle=black,
  title=\textbf{Setence-level Duration Prediction Prompt},
  breakable,
  sharp corners,
  boxrule=0.5pt
]
You are a professional music composer and vocal arranger.

Your task:

1. Analyze the lyrics and the song description below.

2. For each line of lyrics, estimate a reasonable singing duration. Base your estimation jointly on:
\begin{itemize}
    \item The intrinsic characteristics of the line itself (e.g., length, phrasing, complexity)
    \item The overall song attributes;
    \item The structural flow of the song, including instrumental breaks, natural pauses, and transitions;
\end{itemize}

3. Return: Output a complete `.lrc` style list with timestamps.        
\\
\\
Below are the target global song description and lyrics. Please follow the instructions above and return the completed .lrc file directly.
\\
\\
\textbf{Song Description}

This pop rock ballad features a male vocalist delivering an emotional and uplifting melody. The mood is warm and introspective, with a gradually intensifying energy that enhances the song’s heartfelt tone. The singer’s voice is soulful and expressive, using subtle dynamic shifts to convey a sense of comfort and encouragement. 
\\

\textbf{Lyrics}

[This piece is the start of the song.]

[This piece is the intro of the song. The song segment features a spare and gentle piano motif, setting a contemplative and soothing mood ...
]

[This piece is the first verse of the song. The segment features a tender vocal performance accompanied by a steady, melodic piano line ...
]

Hey Jude, don't make it bad, \\
Take a sad song and make it better.\\
Remember to let her into your heart,\\
Then you can start to make it better.\\
...
\\
\\
LRC Prediction:

\end{tcolorbox}

\section{Effect of Post-Training Stages on SegTune} \label{appendix:C}

\newcolumntype{Y}{>{\centering\arraybackslash}X}
\begin{table}[htbp]
\centering
\small
\caption{Performance of SegTune variants on AudioBox-aesthetic and SongEval metrics. SegTune-SFT denotes the model that has undergone both pretraining and supervised fine-tuning (SFT), while SegTune-DPO-$n$ refers to the model further refined via $n$ iterations of Direct Preference Optimization (DPO) starting from the SegTune-SFT checkpoint.}
\label{tab:performance_comparison_segtune_appendixC}

\begin{tabularx}{0.93\textwidth}{@{} >{\centering\arraybackslash}p{1.5cm} *{9}{>{\centering\arraybackslash}p{0.65cm}} @{}}
\toprule
\addlinespace[4pt] 
\multirow{4}{*}{\textbf{Models}}  & \multicolumn{4}{c}{\textbf{AudioBox-aesthetic}} & \multicolumn{5}{c}{\textbf{SongEval}} \\
\addlinespace[4pt] 
\cmidrule(lr){2-5} \cmidrule(lr){6-10}
\addlinespace[4pt] 
&  \multicolumn{1}{c}{\textbf{CE$\uparrow$}} & \multicolumn{1}{c}{\textbf{CU$\uparrow$}} & \multicolumn{1}{c}{\textbf{PC$\uparrow$}} & \multicolumn{1}{c}{\textbf{PQ$\uparrow$}} & \multicolumn{1}{c}{\textbf{Coh$\uparrow$}} & \multicolumn{1}{c}{\textbf{Mem$\uparrow$}} & \multicolumn{1}{c}{\textbf{NVBP$\uparrow$}} & \multicolumn{1}{c}{\textbf{CSS$\uparrow$}} & \multicolumn{1}{c}{\textbf{OM$\uparrow$}} \\
\addlinespace[4pt] 
\midrule

\addlinespace[6pt] 
\textbf{SegTune} &    &  &  &  &  &  &  &  &  \\[2ex]
- SFT &  7.38  & 7.71 & \textbf{6.83} & 8.23 & 3.54 & 3.22 & 3.23 & 3.32 & 3.19\\
\addlinespace[4pt]
- DPO-1 &    7.54&  7.81&  \underline{6.82}&  8.33&  4.00&  3.75&  3.79&  3.81& 3.68\\
\addlinespace[4pt]
- DPO-2 &   \underline{7.63} & \underline{7.85} & 6.80 & \underline{8.36} & \underline{4.25} & \underline{4.06} & \underline{4.09} & \underline{4.08} & \underline{3.97}\\
\addlinespace[4pt] 
- DPO-3 &   \textbf{7.69} & \textbf{7.89} & 6.74 & \textbf{8.39} & \textbf{4.41} & \textbf{4.25} & \textbf{4.29} & \textbf{4.24} & \textbf{4.15}\\

\addlinespace[4pt] 
\bottomrule
\end{tabularx}
\end{table}

\begin{table}[htbp]
\centering
\small
\vspace{1em}
\caption{Performance of SegTune variants on insturction-following metrics.}
\label{tab:dur_predictor_instruction_following_appendixC}

\begin{tabularx}{0.72\textwidth}{>{\centering\arraybackslash}p{1.5cm}  *{4}{>{\centering\arraybackslash}p{1.5cm}} @{}}
\toprule
\addlinespace[6pt] 
\multirow{4}{*}{\textbf{Models}}  & \multicolumn{4}{c}{\textbf{Instruction-following}} \\
\addlinespace[4pt] 
 \cmidrule(lr){2-5} 
\addlinespace[4pt] 
& \textbf{Global Mulan$\uparrow$} & \textbf{Segment Mulan$\uparrow$} & \textbf{Gender$\uparrow$} & \textbf{Age$\uparrow$}  \\
\addlinespace[4pt] 
\midrule
\addlinespace[4pt] 
 \multicolumn{1}{l}{SegTune-SFT} &  0.465 & \textbf{0.378} & \textbf{96.7\%} & \textbf{57\%}  \\
  \addlinespace[4pt] 
 \multicolumn{1}{l}{SegTune-DPO-1}  & 0.456 & 0.361 & 79.1\% & \underline{55\%}  \\
 \addlinespace[4pt] 
 \multicolumn{1}{l}{SegTune-DPO-2}  & \textbf{0.466} & \underline{0.373} & \underline{81.0\%} & 51\%  \\
 \addlinespace[4pt] 
\multicolumn{1}{l}{SegTune-DPO-3}  & \underline{0.465} & 0.373 & 74.3\% & 53\%  \\
 \addlinespace[4pt] 
 
\bottomrule
\end{tabularx}
\end{table}

\end{document}